%% file: Cmpt-EnSc-revision-Dec6-2006.tex
\documentclass[11pt]{article}
\usepackage{amsmath,amsfonts,mathrsfs,latexsym,color,array}

\newcommand{\bR}{{\mathbb{R}}}   
\newcommand{\bref}[1]{(\ref{#1})}
\newcommand{\Proof}{\noindent {\bf Proof:}\ \ }

\newcommand{\R}{\mathbb R}

\newcommand{\B}{{\mathcal B}}
\newcommand{\A}{{\mathcal A}}

\newcommand{\calD}{{\mathcal D}}
\newcommand{\Si}{\Sigma}
\newcommand{\g}{\gamma}
\newcommand{\tg}{{\tilde\gamma}}

\newcommand{\cp}{\frac{4}{n-2}}
\newcommand{\critp}{\frac{n+2}{n-2}}

\newcommand{\di}{\hbox{div}}

\newcommand{\tr}{{\mbox{\rm tr\,}}}

\newcommand{\eop}{\hfill$\Box$}

\newcounter{shownewstuffflag}

\newcommand{\startnewstuff}{\ifnum\value{shownewstuffflag}>0\color{blue}\fi}
\newcommand{\finishnewstuff}{\ifnum\value{shownewstuffflag}>0\color{black}\fi}

\newcounter{oldeq}

\newcounter{mnotecount}[section]

\newcommand{\rmnote}[1]{}

\def\beq{\begin{equation}}
\def\eeq{\end{equation}}

\newtheorem{theorem}{Theorem}

\newtheorem{proposition}{Proposition}

\newtheorem{remark}{Remark}

\begin{document}
\title{The constraint equations for the Einstein-scalar field  system on compact manifolds}

\author{Yvonne Choquet-Bruhat 
\\ University of Paris \and
James Isenberg\thanks{Partially supported by the NSF under Grant
 PHY-0354659 }
\\ University of Oregon  \and
Daniel Pollack\thanks{Partially supported by the NSF under Grant DMS-0305048}
\\ University of Washington}

\date{December 9,  2006}

\maketitle

\abstract{We study the constraint equations for the Einstein-scalar field  system on compact manifolds.
Using the conformal method we reformulate these equations as a determined system of nonlinear partial differential
equations.  By introducing a new conformal invariant, which is sensitive to the presence of the initial data for 
the scalar field, we are able to divide the set of free conformal data into subclasses depending on the possible 
signs for the coefficients of terms in the resulting Einstein-scalar field Lichnerowicz equation.  For many of these subclasses
we determine whether or not a solution exists.  In contrast to other well studied field theories, there are certain
cases, depending on the mean curvature and the potential of the scalar field, for which we are unable to resolve the 
question of existence of a solution. We consider this system in such generality so as to include the vacuum constraint 
equations with an arbitrary cosmological constant, the Yamabe equation and even (all cases of) 
the prescribed scalar curvature problem as special cases.}

\section{Introduction}
\label{se:intro}

While much is understood about constant mean curvature (CMC) solutions of the constraint equations for the vacuum Einstein, Einstein-Maxwell and Einstein-Yang-Mills, as well as a number of other such field theories \cite{CBY, I95, IMaxP}, much less is known about CMC solutions of the Einstein-scalar field constraint equations. This is because when the conformal method is applied in the Einstein-scalar case, the Lichnerowicz equation includes terms of a type which are not seen in these other cases, and which cause some measure of difficulty in the analysis of solvability. In an earlier paper \cite{CBIP}, we show how to overcome some of these difficulties in the asymptotically Euclidean case; here, we work with the Einstein-scalar field constraints on a compact manifold.

For the Einstein vacuum, Einstein-Maxwell, and other cases examined earlier, a complete picture is obtained: We know exactly which sets of CMC conformal data lead to solutions and which do not. This is not true for the Einstein-scalar field constraints: As we see below, there are some sets of conformal data for which it is not known whether the Lichnerowicz equation can be solved. On the other hand, we show that for large sets of Einstein-scalar conformal data, we can determine whether or not a solution exists. 

Interest in Einstein-scalar field theories stems partly from recent attempts to use such theories to explain the observed acceleration of the expansion of the universe \cite{Ren1, Ren2, Ren3, Sa}. Equally important, from the mathematical point of view, is the unique form that the Lichnerowicz equation takes for Einstein-scalar theories.  We use the conformal method to derive the Einstein-scalar field version of the Lichnerowicz equation in \S \ref{eins-scal-sys}, comparing it with the ``usual form" that the Lichnerowicz equation takes. In \S \ref{conflemma}, we state an extension of results concerning conformal transformations of a metric $\gamma$  and the sign of the corresponding scalar curvature $R(\gamma)$; the extension concerns the sign of the quantity $R(\gamma)- |\nabla \psi|^2_\gamma$, where $\psi$ is the scalar field. Using this result, we then catalog in \S \ref{results} those sets of Einstein-scalar conformal data for which either i) a solution to the Einstein-scalar Lichnerowicz equation exists, ii) a solution does not exist, or iii) we have not obtained a conclusive
answer regarding existence, but have partial results.  These results, which appear in Table 1 and Table 2, lead to our main conclusions
which are stated in Theorems \ref{nothm} and  \ref{yesthm} and in the partial results presented in \S \ref{PR} through \S \ref{YCB}.
In \S \ref{proofs} we provide the analysis of the Einstein-scalar Lichnerowicz equation which is required 
to establish these results including the reductions of our problem to various known results for the Yamabe and prescribed
scalar curvature problem.  We conclude in \S \ref{AH} with a discussion of how these results may be extended to allow for 
for asymptotically hyperbolic initial data.

\medskip
\noindent
{\bf Acknowledgements:} The authors would like to thank Helmut Friedrich, David Maxwell and Niall \'O Murchadha for their interest and useful discussions regarding the work presented here.    The authors would like to thank the Isaac Newton Institute of Mathematical Sciences in Cambridge, England  
for  providing an excellent research environment  during the program on Global Problems in 
Mathematical Relativity during Autumn, 2005, and again in October 2006, where much of  this research was carried out.

\section{The Constraint Equations for the Einstein-Scalar Field Theories}
\label{eins-scal-sys}

The field variables for an Einstein-scalar field theory consist of a spacetime metric $g$ and a real-valued\footnote{For many purposes, it is useful to consider complex-valued scalar fields. For convenience, we restrict to the real case here.} scalar field $\Psi$, both specified on an $(n+1)$-dimensional spacetime manifold $M$. 
For convenience here we assume that all variables, and subsequent choices of initial data, are taken to be smooth. 
However, in practice our results hold both in the smooth category and when working in function spaces of 
finite partial regularity, as is discussed at the beginning of \S \ref{proofs} below.
A particular Einstein-scalar field theory is specified by the choice of an action principle taking the form\footnote{We use MTW \cite{MTW} conventions here for the signature of the metric, the definition of the curvatures, and the assignment of indices. We note in particular that Greek indices refer to spacetime, Latin indices refer to space, and the index $ ``\perp"$ refers to the unit timelike  vector field normal to the leaves of a $n+1$ foliation of spacetime. Note also that we have also chosen units so that 
$8\pi G=1=c$.}
\begin{equation}
\mathcal{S}(g,\Psi) = \int_M [R(g) - \frac{1}{2} |\nabla \Psi|^2_g  - V(\Psi)] d\eta_g, 
\end{equation}
where $R(g)$ is the scalar curvature of $g$, $d\eta_g$ is its volume element, $ |\nabla \Psi|^2_g$ is the squared pseudo-norm of the  spacetime gradient of $\Psi$ taken with respect to the metric $g$ and the potential $V(\cdot)$ is a given smooth function of 
a real variable. (For the massive Klein-Gordon field theory, 
$V(\Psi) = \frac{1}{2}m^2 \Psi^2$, for a constant $m$ specifying the mass.) 
Note that we do not require that  $V$ should vanish at $\Psi\equiv 0$. Thus, for appropriate choices of $V$, this gives rise
to the presence of an arbitrary cosmological constant.  We may also, by making a particular choice of $V(\Psi)$ independent
of $\Psi$, obtain the prescribed scalar curvature problem as a special case of the resulting constraint equations.
Varying this Einstein-scalar field action with respect to the fields $g$ and $\Psi$, we obtain the Einstein-scalar field equations 
corresponding to the given (and at this stage arbitrary) choice of  $V(\Psi)$:
\begin{equation} \label{Ein}
G_{\alpha \beta} = T_{\alpha \beta} = \nabla_\alpha \Psi \nabla_\beta \Psi -\frac{1}{2}g_{\alpha \beta} \nabla_\mu \Psi \nabla^\mu \Psi -g_{\alpha \beta} V,
\end{equation}
\begin{equation} \label{Scal}
\nabla_\mu \nabla^\mu \Psi = \frac{dV}{ d\Psi}.
\end{equation}
(where $G_{\alpha \beta}$ is the Einstein curvature tensor).

The coupling of a scalar field to the Einstein gravitational field theory does not add any new constraint equations to the theory. We have the usual Hamiltonian and momentum constraints (the $G_{\perp \perp}$ and $G_{\perp a}$ equations ), but  with added scalar field source terms. Writing these out in terms of the $n+1$ decomposition fields on an $n$-dimensional spacelike hypersurface $\Sigma$ $\{{\bar \gamma}$ (the spatial metric), ${\bar K}$ (the second fundamental form, or extrinsic curvature), ${\bar \psi}$  (the scalar field restricted to 
$\Sigma$), ${\bar \pi}$ (the normalized time derivative of $\Psi$ restricted to $\Sigma$)\} we have
\begin{eqnarray}
\label{ham} 
R({\bar \gamma}) -|{\bar K}|^2_{\bar \gamma} +(\tr {\bar K})^2 &=& {\bar \pi}^2 + |\nabla {\bar \psi}|^2_{\bar \gamma}  +2V({\bar \psi})\\
\label{mom}
{\di_{\bar\gamma} \bar K}-\nabla (\tr{\bar  K}) &=& -{\bar \pi} \nabla {\bar \psi},
\end{eqnarray}
where all derivatives and norms are taken with respect to the metric $\bar\gamma$ on $\Sigma$.
These  constraints are to be solved for the Cauchy data $({\bar \gamma}, {\bar K}, {\bar \psi}, {\bar \pi})$ on a chosen $n$-dimensional manifold $\Sigma$. Local well-posedness theorems \cite{FB52} then guarantee that there is an $(n+1)$-dimensional spacetime solution $(M=\Sigma \times \R,g,\Psi)$ of (\ref{Ein})-(\ref{Scal}) which is consistent with the given Cauchy data.

The idea of the conformal method is to recast the constraint equations (\ref{ham})-(\ref{mom}) into a form which is more amenable to analysis,  by splitting the Cauchy data into (i) the ``conformal data", which one can choose freely, and (ii) the ``determined data", which is determined by solving the recast constraints. For the gravitational data, one achieves an optimal form\footnote{There are other ``optimal" splits of the gravitational data which have various advantages (along with disadvantages) relative to the one discussed here.  Some of them are reviewed in \cite{BI04}; see also \cite{PY03}. However for the purposes of the present work, the differences among the various methods are not important.}  via the decomposition of the covariant 2-tensors
\begin{equation} \label{confmetric}
{\bar \gamma} = \phi^{\frac{4}{n-2}} \gamma
\end{equation} 
\begin{equation} \label{confK}
{\bar K} = \phi^{-2} (\sigma + {\cal D}W) + \frac{\tau}{n} \phi^\cp \gamma
\end{equation}
where the conformal data consists of a Riemannian metric $\gamma=\gamma_{ab}$, a symmetric tensor 
$\sigma=\sigma_{ab}$ which is divergence-free and trace-free with respect to $\gamma$ (so that $\sigma$ is
what is commonly referred to as a TT-tensor) and a scalar $\tau$ representing the mean curvature
of the Cauchy surface $\Sigma$ in the resulting spacetime; while the determined data consists of the positive function $\phi$ and the vector field $ W=W^a$. Here the operator $\cal D$ is  the conformal Killing operator
relative to $\gamma$,  defined by $({\cal D}W)_{ab} := \nabla_a W_b+ \nabla_b W_a -\frac{2}{n} \gamma_{ab}\nabla_m W^m$, 
where $\nabla$ is the covariant derivative for the metric $\g$.
The  kernel of $\cal D$  consists of conformal Killing fields. This particular form of the decomposition has the virtue that, since 
\begin{equation}
\label{scalcurv}
R({\bar \gamma}) = -\phi^{-\critp} \left(\frac{4(n-1)}{n-2} \Delta_{\gamma} \phi - R(\gamma)\phi\right)
\end{equation}
and
\begin{equation} 
\nabla^a_{\bar\gamma} (\phi^{-2} B_{ab}) = \phi^{-\frac{2n}{n-2}}\nabla^a_{\gamma} ( B_{ab})
\end{equation}
for any trace-free tensor $B$, 
it avoids the appearance of $\nabla\phi$  terms and (if $\tau$ is constant) $\phi$ terms in the momentum equation,  and it avoids $|\nabla \phi|^2_\gamma$ terms in the Hamiltonian equation. 

In determining how to decompose the scalar field initial data $({\bar \psi}, {\bar\pi})$, our goal is to preserve the three virtues  just mentioned.  In addition, since the gravitational decomposition has introduced (locally) $n+1$ unknown functions ($\phi$ and the components of $W$) as determined data, and since there are essentially $n+1$ constraint equations to determine this data, in the scalar field decomposition we must refrain from introducing  any new unknowns. One readily finds that the only scalar field decomposition which satisfies these objectives is the following:
\begin{equation} \label{psi}
{\bar \psi} = \psi
\end{equation}
\begin{equation} \label{pi}
{\bar\pi} = \phi ^{-\frac{2n}{n-2}} \pi.
\end{equation}
(This corresponds to the ``York scaling" of the initial data \cite{Y72, CBY} which provides for the decomposition demonstrated here.)
Combining these decompositions of the gravitational and the scalar field data, we write out the conformal form of the constraint equations as follows:
\begin{equation} \label{lich}
\begin{array}{ccc}
\Delta_{\gamma}\phi - \frac{n-2}{4(n-1)}\left(R(\gamma)-|\nabla\psi|^2_{\gamma}\right) \phi
&+& \frac{n-2}{4(n-1)}\left(|\sigma + {\cal D} W|^2_{\gamma}+\pi^2\right) \phi^{-\frac{3n-2}{n-2}}\\
&-&\frac{n-2}{4(n-1)}\left(\frac{n-1}{n} \tau^2 -4V(\psi)\right)\phi^\frac{n+2}{n-2} 
= 0.
\end{array}
\end{equation}
\begin{equation} \label{confmom}
\di_{\gamma}({\cal D} W) = \frac{n-1}{n}\phi^{\frac{2n}{n-2}}\nabla \tau - \pi\nabla\psi.
\end{equation}
As claimed, this system avoids derivatives of $\phi$ apart from the Laplacian term, and equation (\ref{confmom}) involves no terms containing $\phi$ at all if $\tau$ is constant.   To facilitate our subsequent
treatment of (\ref{lich}) we make the following definitions.  We set $c_n = \frac{n-2}{4(n-1)}$,  
$p_n=\frac{2n}{n-2}$ and let
\[
 {\cal R}_{\gamma, \psi}=c_n\left(R(\gamma)-|\nabla{\psi}|^2_{\gamma}\right), \qquad
 {\cal A}_{\gamma, W, \pi}=c_n\left(|\sigma + {\cal D} W|^2_{\gamma}+\pi^2\right)
 \]
 and
 \[
{\cal B}_{\tau, \psi}=c_n\left(\frac{n-1}{n} \tau^2 -4V(\psi) \right).
\]
We may then rewrite the Lichnerowicz equation (\ref{lich}) for the Einstein-scalar conformal data 
$(\gamma,  \sigma, \tau, \psi, \pi)$ with a given vector field $W$ satisfying  (\ref{confmom}) as
\begin{equation} \label{shortlich}
\Delta_{\gamma}\phi -  {\cal R}_{\gamma, \psi}\,\phi + {\cal A}_{\gamma, W, \pi}\,\phi^{-\frac{3n-2}{n-2}}
-  {\cal B}_{\tau, \psi}\,\phi^{\frac{n+2}{n-2}}=0.
\end{equation}
Our main interest here is the determination of which choices of the conformal data $(\gamma, \sigma, \tau, \psi, \pi)$ permit one to solve the conformal constraint equations (\ref{lich})-(\ref{confmom}) for the determined data $(\phi, W)$ and which do not. (Recall that in this paper, we always presume that the $n$-manifold $\Sigma$ is closed; i.e., compact without boundary.)  We are particularly interested in handling the special complications which the addition of the scalar field causes, in comparison with the addition of other fields such as the Maxwell electromagnetic field. These complications arise exclusively in equation (\ref{lich}), so our discussion here focuses on this equation. 
(See \S \ref{results} for the application of our analysis to the full conformal constraint equations (\ref{lich})-(\ref{confmom}) in 
the uncoupled case, when $\tau$ is constant.)  To aid the discussion, we include the version of this equation which arises for the Einstein-Maxwell field theory in 3+1 dimensions:
\begin{equation} \label{LichMax}
\Delta \phi - \frac18 R(\gamma)\,\phi + \frac18 ( |\sigma + {\cal D} W|^2_{\gamma})\phi^{-7} +
\frac18(|E|^2_\gamma+ |B|^2_\gamma) \phi ^{-3} -   \frac{1}{12} \tau^2 \phi ^5 =0.
\end{equation}
(Here the vector fields $E$ and $B$, representing the electric and magnetic fields respectively, are included in the set of conformal data. For the Einstein-Maxwell system we have the additional constraints that these
vector fields are divergence free.) We note that the form this equation takes for the Einstein-Yang-Mills system is essentially the same. 

There are a variety of differences between (\ref{lich}) and (\ref{LichMax}).
One important distinction is the introduction of the potential
$V(\psi)$ in the coefficient of the $\phi^{\frac{n+2}{n-2}}$ term.  For many choices of potential and scalar field, this results in a coefficient 
which does not maintain a constant sign on $\Sigma$.  The methods currently available for solving  (\ref{lich}) make crucial use
of the signs of the coefficients (see \S\ref{proofs}). As a result we are best able to resolve the question of the existence of
solutions under the assumption that the coefficient ${\cal B}_{\tau, \psi}$ does not change sign on $\Sigma$; however
we also have partial results (see \S\ref{YCB}) when ${\cal B}_{\tau, \psi}$ changes sign.
Also noteworthy, as far as the analysis of solubility is concerned, is the replacement of $c_n R(\gamma)\phi$ as the linear term  in (\ref{LichMax})  (which is also the linear, zero-order term in the vacuum Lichnerowicz equation) by the term $c_n(R(\gamma) - |\nabla\psi|^2_\gamma)\phi$ in (\ref{lich}).  This replacement is 
important because while there are well-known criteria for ascertaining the possible signs of the scalar curvature function with a given conformal class of metrics, such results are not in the literature for the quantity $c_n(R(\gamma) - |\nabla\psi|^2_\gamma)$.  We remedy this in the next section.

\section{The Yamabe-scalar field conformal invariant  and the sign of $R(\gamma) - |\nabla\psi|^2_\gamma$}
\label{conflemma}

In determining whether or not the Lichnerowicz equation, for a given set of vacuum Einstein conformal data $(\gamma, \sigma, \tau)$, can be solved for a positive function $\phi$, one relies strongly on two results: (i) the conformal covariance of the Lichnerowicz equation (specifically, one finds that  there exists a positive solution $\phi$ for the conformal data $(\gamma, \sigma, \tau)$ if and only if there exists a positive solution $\tilde{\phi}$
for the conformal data $(\theta^{\frac{4}{n-2}} \gamma, \theta ^{-2}\sigma, \tau)$ for any positive function $\theta$); and (ii)  the fact that every Riemannian metric $\gamma$ can be conformally transformed in such a way that the scalar curvature of the transformed metric is either everywhere positive, everywhere negative, or everywhere zero. 

One readily shows that for the Einstein-scalar Lichnerowicz equation \bref{lich}, conformal covariance still holds in the appropriately modified sense (see Proposition \ref{confinv} in \S \ref{proofs} below).   In order to have a useful analog of the second result we proceed
as follows.  
\begin{remark}
The discussion here is exactly analogous to the definitions and results for the usual Yamabe 
conformal invariant (which are recovered by setting $\psi=0$ below).  We have included the 
proof of Proposition \ref{tfae} below for completeness; however the reader is referred to 
\cite{LP, Yam, Tr, Au1, Au2, Sc-n} 
for the original formulations in the absence of a scalar field.
\end{remark}

Recall that on a Riemannian manifold $(\Sigma, \gamma)$, the conformal Laplace
operator $L_\gamma$ acting on a   smooth 
function $u$ is defined by $L_\gamma u = \Delta_\gamma u - c_n R(\gamma) u$. 
Given a smooth function
$\psi:\Sigma \rightarrow \bR$, we define the {\em conformal scalar field Laplace operator} $L_{\gamma, \psi}$ by 
\begin{equation}
L_{\gamma, \psi}u = \Delta_\gamma u - c_n \left(R(\gamma) - |\nabla \psi|^2_\gamma\right) u.
\end{equation}
If ${\tilde\gamma} = \theta^{\frac{4}{n-2}}\gamma$ for some $\theta>0$ then, using the conformal covariance property
of the conformal Laplacian, one may easily show that $L_{\gamma, \psi}$ satisfies the same conformal 
covariance, namely
\begin{equation}
\label{confcovar}
 L_{{\tilde\gamma}, \psi}u =\theta^{-\frac{n+2}{n-2}}L_{\gamma, \psi}(\theta u)
\end{equation}
for any smooth function $u$.  Applying this when $u\equiv 1$ we obtain the formula exhibiting  how 
$c_n \left(R(\gamma) - |\nabla \psi|^2_\gamma\right)$ changes under a conformal change of metric
\[
c_n \left(R(\tilde\gamma) - |\nabla \psi|^2_{\tilde\gamma}\right) = -\theta^{-\frac{n+2}{n-2}}L_{\gamma, \psi}(\theta).
\]
One immediately recognizes the similarity between this and the  corresponding formula for $R(\tilde\gamma)$ alone.
We define the {\em conformal-scalar Dirichlet} energy of $u$ by
\begin{eqnarray*}
E_{\gamma, \psi}(u) &=& c_n^{-1} \int_\Sigma -u  L_{\gamma, \psi}u \, d\eta_\gamma\\
&=& c_n^{-1} \int_\Sigma [|\nabla u|^2_\gamma + c_n\left(R(\gamma) - |\nabla \psi|^2_\gamma\right) u^2] \,d\eta_\gamma,
\end{eqnarray*}
and the {\em conformal-scalar field Sobolev quotient} by
\begin{eqnarray*}
Q_{\gamma, \psi}(u) &=& \frac{{\displaystyle E_{\gamma, \psi}(u)}}{{\displaystyle\|u\|^2_{p_n}}}\\
&=&\frac{{\displaystyle c_n^{-1} \int_\Sigma [|\nabla u|^2_\gamma + c_n\left(R(\gamma) - |\nabla \psi|^2_\gamma\right) u^2] \,d\eta_\gamma}}{{\displaystyle\left( \int_\Sigma u^{\frac{2n}{n-2}}\,d\eta_\gamma\right)^{\frac{n-2}{n}} }}. 
\end{eqnarray*}
Using (\ref{confcovar}) ones easily sees that
\begin{equation}
\label{qconfinv}
Q_{\tilde\gamma, \psi}(u) = Q_{\gamma, \psi}(\theta u).  
\end{equation}
We denote the conformal class of the metric $\gamma$ by
\[
[\gamma] = \{\tilde\gamma = \theta^{\frac{4}{n-2}}\gamma: \theta\in C^{\infty}(\Sigma), \theta>0\}.
\]
The Yamabe-scalar field conformal invariant is then defined by
\begin{equation}
\label{yamscal}
{\mathcal Y}_{\psi}([\gamma]) = \inf_{u\in H^1(\Sigma)} Q_{\gamma, \psi}(u) = \inf_{u\in H^1(\Sigma)}
\frac{{\displaystyle c_n^{-1} \int_\Sigma [|\nabla u|^2_\gamma + c_n\left(R(\gamma) - |\nabla \psi|^2_\gamma\right) u^2] \,d\eta_\gamma}}{{\displaystyle\left( \int_\Sigma u^{\frac{2n}{n-2}}\,d\eta_\gamma\right)^{\frac{n-2}{n}} }}.
\end{equation}
By (\ref{qconfinv}) it is immediate that ${\mathcal Y}_{\psi}([\gamma])$ is independent of the 
choice of background metric in the conformal class used to define it, and is therefore an invariant of the 
conformal class.  Using H\"older's inequality we observe that $|\int_\Sigma (R(\gamma) - |\nabla \psi|^2_\gamma) u^2] \,d\eta_\gamma | \leq c \|u\|_{p_n}^2$ for some constant $c$ independent of $u$.  This shows that
${\mathcal Y}_{\psi}([\gamma])>-\infty$. 

\begin{proposition} \label{tfae} 
The following conditions are equivalent:
\begin {itemize}
\item[(i)] ${\mathcal Y}_{\psi}([\gamma])>0$  (respectively $=0, <0$).
\item[(ii)] There exists a metric $\tilde\gamma\in [\gamma]$ which satisfies
 $(R(\tilde\gamma) - |\tilde\nabla \psi|^2_{\tilde\gamma})>0$ everywhere on $\Sigma$
 (respectively $=0, <0$).
\item[(iii)] For any metric $\tilde\gamma\in [\gamma]$, the first eigenvalue, $\lambda_1$,
of the self-adjoint, elliptic operator $-L_{{\tilde\gamma}, \psi}$ is positive (respectively zero, negative).
\end{itemize}
\end{proposition}

\Proof
We work in the positive case and follow each implication with  remarks as to what modifications, if any,
are required in the zero or negative cases.

$(iii)\Rightarrow (ii)$: Suppose $\lambda_1(-L_{\gamma, \psi})>0$.  We may take $v>0$ to be a
positive eigenfunction for  $-L_{\gamma, \psi}$ with eigenvalue $\lambda_1$, so that 
\[
-L_{\gamma, \psi}v= \lambda_1 v>0.
\]
Let $\tg = v^\frac{4}{n-2}\g$.  Applying (\ref{scalcurv}) we see that
\begin{eqnarray*}
R(\tilde\gamma) - |\tilde\nabla \psi|^2_{\tilde\gamma} 
&=& -c_n^{-1} v^{-\frac{n+2}{n-2}}\left(\Delta_\g v- c_nR_\g v\right)- |\tilde\nabla \psi|^2_{\tilde\gamma}\\
&=& -c_n^{-1} v^{-\frac{n+2}{n-2}}\left(\Delta_\g v- c_nR_\g v\right)- v^{-\frac{4}{n-2}}|\nabla \psi|^2_{\gamma}\\
&=& -c_n^{-1} v^{-\frac{n+2}{n-2}}\left(\Delta_\g v- c_n(R_\g v- |\nabla \psi|^2_{\gamma}v)\right)\\
&=& c_n^{-1} v^{-\frac{n+2}{n-2}}\left(-L_{\gamma, \psi}v\right)=c_n^{-1} v^{-\frac{n+2}{n-2}} \lambda_1 v>0.
\end{eqnarray*}
This verifies that $(ii)$ holds.

Note that this same argument verifies the desired implication in the zero and negative cases as well.

$(ii)\Rightarrow (i)$: Let $\tilde\gamma\in [\gamma]$ satisfy 
\beq
\label{curvbnd}
(R(\tilde\gamma) - |\tilde\nabla \psi|^2_{\tilde\gamma})>0.
\eeq 
For any function $u\in H^1(\Sigma)$ we compute its energy with respect to the background metric $\tg$.
\[
E_{\tg, \psi}(u) = c_n^{-1} \int_\Sigma [|\tilde\nabla u|^2_{\tg} + c_n\left(R(\tg) - |\tilde\nabla \psi|^2_{\tg}\right) u^2] \,d\eta_{\tg}.
\]
By (\ref{curvbnd}) we see that the energy bounds the square of the $H^1$-norm of $u$,
\[
c\|u\|^2_{1,2} \leq E_{\tg, \psi}(u)
\]
for some constant $c>0$. Recall that the Sobolev embedding theorem asserts that 
$H^1(\Si)\hookrightarrow L^{p_n}(\Si)$ is a bounded linear operator. Thus, there is a 
positive constant $c_1$ such that 
\[
c_1\|u\|_{p_n}^{2}  \leq c\|u\|^2_{1,2} \leq E_{\tg, \psi}(u).
\]
From this we conclude that the conformal scalar field Sobolev quotient of $u$ satisfies
\[
Q_{\tg, \psi}(u) = \frac{E_{\tg, \psi}(u)}{\|u\|_{p_n}^{2} }\geq c_1 >0.
\]
Since $u\in H^1(\Si)$ is arbitrary, we conclude that  ${\mathcal Y}_{\psi}([\gamma])>0$, verifying
$(i)$.

In the negative case the desired assertion is established instead by choosing a constant test function, 
$u=1$. It immediately follows that  $Q_{\tg, \psi}(1) <0$ and hence ${\mathcal Y}_{\psi}([\gamma])<0$.  

In the zero case, since $Q_{\tg, \psi}(1) =0$ we conclude that  ${\mathcal Y}_{\psi}([\gamma])\leq 0$.
If here it were the case that ${\mathcal Y}_{\psi}([\gamma])< 0$ then there would exist a $u$ 
with $Q_{\tg, \psi}(u) <0$. This is a contradiction since $R(\tilde\gamma) - |\tilde\nabla \psi|^2_{\tilde\gamma}=0$
implies that $Q_{\tg, \psi}(u) \geq0$ for all $u$.
We therefore conclude that ${\mathcal Y}_{\psi}([\gamma])=0$ as required. 

$(i)\Rightarrow (iii)$:  Suppose that ${\mathcal Y}_{\psi}([\gamma])>0$.  Given any $\tg\in [\g]$, let 
$v>0$ be the first eigenfunction of $-L_{{\tilde\gamma}, \psi}$ normalized to have $L^{p_n}$-norm 
equal to one.  We then have 
\[
E_{\tg, \psi}(v)\geq {\mathcal Y}_{\psi}([\gamma])>0.
\]
On the other hand
\begin{eqnarray*}
E_{\tg, \psi}(v) &=& c_n^{-1} \int_\Sigma -v  L_{\tg, \psi}v \, d\eta_\tg\\
&=& c_n^{-1} \lambda_1 \int_\Sigma v^2 \, d\eta_\tg
\end{eqnarray*}
where $\lambda_1$ is the first eigenvalue.  From this we conclude that   $\lambda_1>0$ as desired. 

This argument shows that $\lambda_1(-L_{{\tilde\gamma}, \psi})\geq 0$ in the case of a 
zero Yamabe-scalar field conformal invariant.  There we must have $\lambda_1(-L_{{\tilde\gamma}, \psi})=0$
in this case, since otherwise we would have a contradiction with the implication $(iii)\Rightarrow (i)$
in the positive case.

For the negative case, we may choose a $v\in H^1(\Sigma)$ for which $Q_{\tg, \psi}(v)<0$ and normalize the volume of $\tg$ to be one.  Then by H\"older's inequality ${\|v\|_{p_n}^2}\leq {\|v\|_2^2}$. Since
\[
\lambda_1(-L_{{\tilde\gamma}, \psi}) = \inf_{u\in H^1(\Sigma)} \frac{E_{\tg, \psi}(u)}{\|u\|_2^2}
\]
we see that
\[
\lambda_1(-L_{{\tilde\gamma}, \psi}) \leq \frac{E_{\tg, \psi}(v)}{\|v\|_2^2}= 
\frac{E_{\tg, \psi}(v)}{\|v\|_{p_n}^2}  
\frac{\|v\|_{p_n}^2}{\|v\|_2^2}
\leq \frac{E_{\tg, \psi}(v)}{\|v\|_{p_n}^2}
= Q_{\tg, \psi}(v)<0.
\]
which establishes $(iii)$ in this case as well. 
\eop

Using the sign of the Yamabe-scalar field conformal invariant we may partition the set of pairs $([\gamma], \psi)$
consisting of a conformal class of Riemannian metrics and a scalar field on $\Sigma$,  into three classes
which we label $\mathcal{Y}^+,  \mathcal{Y}^0,  \mathcal{Y}^-$, and refer to as the positive, zero, and negative 
Yamabe-scalar field classes on $\Sigma$. 
For a given set of conformal data $(\g, \sigma, \tau, \psi, \pi)$ for the Einstein-scalar field constraint 
equations on $\Sigma$, the Yamabe-scalar field class plays an important role in determining whether the 
Einstein-scalar Lichnerowicz equation can be solved. This is because, as a consequence of Proposition \ref{tfae}, 
we may always perform a conformal transformation on the conformal data  $(\g, \sigma, \tau, \psi, \pi) \rightarrow  
(\theta^{\frac{4}{n-2}} \g, \theta^{-2}\sigma, \tau,  \psi, \theta^{-\frac{2n}{n-2}}\pi)$ in such a way that 
$ {\cal R}_{\gamma, \psi}$ (the zero order coefficient of $\phi$ in the Einstein-scalar Lichnerowicz equation \bref{lich}) 
is either greater than zero everywhere, equal to zero everywhere, or less than zero everywhere. It is easier to analyze 
solvability of \bref{lich} for this transformed data. Then, since it follows as a consequence of the conformal 
covariance of the Einstein-scalar Lichnerowicz equation (see Proposition \ref{confinv} below) 
that the equation admits a positive solution for the original data 
if and only if it admits a positive solution for the transformed data, the consequences of 
this easier analysis holds for the original data. 

\section{Main results}
\label{results}

\subsection{Solving the Momentum constraint}

The operator $\di_{\gamma}\circ{\cal D}$ appearing in (\ref{confmom}) is a second order, self-adjoint, linear, elliptic operator 
whose kernel consists of the space of conformal Killing vector fields on $(\Sigma, \g)$.  It follows that for a given set of 
functions $(\phi, \tau, \psi, \pi)$ we may solve (\ref{confmom})  provided $\frac{n-1}{n}\phi^{\frac{2n}{n-2}}\nabla \tau - \pi\nabla\psi$
is orthogonal to this space.  The resulting solution is unique if and only if the space of conformal Killing vector fields on 
$(\Sigma, \g)$ is empty.  

When the mean curvature function $\tau$ is non-constant we observe that, since the conformal factor $\phi$ appears in 
 (\ref{confmom}), the system (\ref{lich})-(\ref{confmom}) is truly coupled.  Due to the presence of the critical exponent $\frac{2n}{n-2}$ 
 in  (\ref{confmom}), even in the vacuum case, we do not know how to address solvability of this coupled 
 system in general.  Hence at present, in order to formulate a general existence result for the full set of constraint equations it is necessary
to assume that we are in the constant mean curvature setting and we may therefore first solve  (\ref{confmom}), under 
 the restrictions given above, for the vector field $W$ and then solve the conformal form of the Hamiltonian constraint 
(\ref{lich})  for  $\phi$.

\subsection{Solving the Hamiltonian constraint}

We recall from earlier work \cite{I95} that in determining which sets of vacuum CMC conformal data $(\g, \sigma, \tau)$ permit the vacuum Lichnerowicz equation to be solved and which do not, we distinguish twelve classes of data, depending on (i) the Yamabe class of the metric, (ii) whether the function $|\sigma|^2$ is identically zero or not, and (iii) whether the constant $\tau$ is zero or not. 

For the Einstein-scalar case, the classification of the data is more complicated, primarily because there are more relevant possibilities for the signs of the coefficients in \bref{lich}. In particular, we need to distinguish six different possibilities for the quantity ${\cal B}_{\tau, \psi}=c_n\left(\frac{n-1}{n} \tau^2 -4V(\psi)\right)$, which in general is non-constant even in the CMC setting.
These categories correspond to whether ${\cal B}_{\tau, \psi}$ is strictly positive, greater than or equal to zero, identically zero,
less than or equal to zero, strictly negative, or of indeterminate sign.
Combining these six categories with the three Yamabe scalar field classes $\{ \mathcal{Y}^+, \mathcal{Y} ^0, \mathcal{Y} ^-\}$ and the two classes
 $\{\A_{\g, W, \pi} \equiv 0, \A_{\g, W, \pi}\not \equiv 0\}$ which reflect whether the coefficient
 \footnote{The coefficient $c_n\left(|\sigma + {\cal D} W|^2_{\gamma}+\pi^2\right)$ involves the determined data field $W$ as 
 well as conformal data. It follows from analysis of the momentum constraint \bref{confmom}, however, that $c_n\left(|\sigma + {\cal D} W|^2_{\gamma}+\pi^2\right)$ vanishes if and only if $|\sigma|_\g^2 + \pi^2$ vanishes, and this quantity depends entirely on the conformal data.} 
$\A_{\g, W, \pi}= c_n\left(|\sigma + {\cal D} W|^2_{\gamma}+\pi^2\right)$ 
of the $\phi^{-\frac{3n-2}{n-2}}$  term vanishes identically or not, we have a total of thirty-six classes to consider. 

According to this division, there are six categories which occur when 
${\cal B}_{\tau, \psi}=c_n\left(\frac{n-1}{n} \tau^2 -4V(\psi)\right)$ changes sign.  We have not determined whether we can solve the
Einstein-scalar field constraints in these cases.  Note that the above division is not ``uniform". 
For most admissible choices of the potential $V(\psi)$, and for most choices of the conformal data component $\psi$, ${\cal B}_{\tau, \psi}$ 
changes sign on $\Sigma$.  However, for a number of the potentials $V$ which arise in physics, there are reasonable restrictions on 
the choice of the conformal data which results in a fixed sign for ${\cal B}_{\tau, \psi}$ .
For example, in the case of the Einstein massive Klein-Gordon system, $V(\psi) =\frac{1}{2}m^2\psi^2$.
We may ensure that  ${\cal B}_{\tau, \psi}$ has a fixed sign by assuming a corresponding inequality
on the mean curvature function $\tau$ in terms of the mass and the minimum or maximum of the specified value of the scalar field $\psi$
on $\Sigma$.  In all but six of the thirty cases in which ${\cal B}_{\tau, \psi}$ has a sign we determine, in \S \ref{proofs}, 
whether the corresponding Einstein-scalar field Lichnerowicz equation can be solved (for a positive function $\phi$) or not. 
In two of the remaining cases, we obtain necessary and sufficient conditions; for the remaining four cases we exhibit sufficient
conditions for the existence of a positive solution.

We collect the results of the analysis which we make in \S\ref{proofs}  in the following two tables, where ``Y" indicates that 
the Lichnerowicz equation can be solved for that class of conformal data, ``N" indicates that the corresponding 
Lichnerowicz equation has no positive solution,  ``PR" indicates that we have partial results and 
``NR" indicates that for this class of initial data we have no results indicating existence or non-existence.

\bigskip
\begin{tabular}{|c||c|c|c|c|c|c|}
\hline
&{\tiny $\B_{\tau,\psi}$ changes sign}&{\tiny$\B_{\tau,\psi}< 0$}&{\tiny$\B_{\tau,\psi}\leq 0$}&{\tiny$\B_{\tau,\psi}\equiv 0$}&{\tiny$\B_{\tau,\psi}\geq 0$}&{\tiny$\B_{\tau,\psi}>0$}\\
\hline
\hline
${\mathcal Y}_{\psi}([\gamma])<0$&NR&N&N&N&PR&Y\\
\hline
${\mathcal Y}_{\psi}([\gamma])=0$&NR&N&N&Y&N&N\\
\hline
${\mathcal Y}_{\psi}([\gamma])>0$&PR&PR&PR&N&N&N\\
\hline
\multicolumn{6}{c}{}\\
\multicolumn{6}{c}{{\bf Table 1}: Results for $\A_{\g, W, \pi}\equiv 0$.} \\
\end{tabular}
\bigskip

\bigskip
\begin{tabular}{|c||c|c|c|c|c|c|}
\hline
&{\tiny$\B_{\tau,\psi}$ changes sign}&{\tiny$\B_{\tau,\psi}< 0$}&{\tiny$\B_{\tau,\psi}\leq 0$}&{\tiny$\B_{\tau,\psi}\equiv 0$}&{\tiny$\B_{\tau,\psi}\geq 0$}&{\tiny$\B_{\tau,\psi}>0$}\\
\hline
\hline
${\mathcal Y}_{\psi}([\gamma])<0$&NR&N&N&N&PR&Y\\
\hline
${\mathcal Y}_{\psi}([\gamma])=0$&NR&N&N&N&Y&Y\\
\hline
${\mathcal Y}_{\psi}([\gamma])>0$&PR&PR&PR&Y&Y&Y\\
\hline
\multicolumn{6}{c}{}\\
\multicolumn{6}{c}{{\bf Table 2}: Results for $\A_{\g, W, \pi}\not\equiv 0$.} \\
\end{tabular}
\bigskip

\subsection{The main theorems}

Using Tables 1 and 2, we state the following theorems.
The first theorem establishes those situations in which we can prove that no solution to the Einstein-scalar field constraint equations exists.
\begin{theorem}
\label{nothm}
Assume that we are given a compact manifold $\Sigma$, and conformal data $(\g, \sigma, \tau, \psi, \pi)$ on $\Sigma$, with $\tau$ constant. 
If the relevant entry
\footnote{As we have already observed (in footnote 4) we may take $W\equiv 0$ in deciding which Table to consult. Therefore our non-existence
results are valid regardless of our ability to satisfy the momentum constraint, and therefore regardless  of the presence
of conformal Killing fields.}
in Table 1 or Table 2 is {\rm N}, then there is no solution to the Einstein-scalar field constraint equations with respect to 
any metric in the conformal class of $\g$ and with $(\psi, \pi)$ as the initial data for the scalar field.
\end{theorem}
The second theorem establishes those cases in which we can prove that  
a solution to the Einstein-scalar field constraint equations exists.
\begin{theorem}
\label{yesthm}
Assume that we are given a compact manifold $\Sigma$, and conformal data $(\g, \sigma, \tau, \psi, \pi)$ on $\Sigma$, with $\tau$ constant.  
If either $(\Sigma, \g)$ admits no conformal Killing vector fields, or $\pi\nabla\psi$ is orthogonal 
to the space of conformal Killing vector fields, and the relevant entry in Table 1 or Table 2 is {\rm Y}, then we may
find a positive function $\phi$ and a vector field $W$ so that the conformal reconstructed data 
$({\bar \g}, {\bar K}, {\bar \psi}, {\bar \pi})$, defined in equations \bref{confmetric}, \bref{confK}, \bref{psi} and \bref{pi},   
satisfy the Einstein-scalar field constraint equations \bref{ham}-\bref{mom}.  In the case that there are no conformal 
Killing fields, the resulting solution is unique provided that any one of $ {\cal R}_{\gamma, \psi}, {\cal A}_{\gamma, W, \pi}$ or  
${\cal B}_{\tau, \psi}$ are not identically zero. In the absence of conformal Killing fields for the trivial case when  
$ {\cal R}_{\gamma, \psi}, {\cal A}_{\gamma, W, \pi}$ and ${\cal B}_{\tau, \psi}$ are  identically zero there is a natural scale invariance
in the solution.
\end{theorem}
For the remaining cases, in which we presently have partial results, we state and prove sufficient conditions for existence in 
\S \ref{PR} through \S \ref{YCB} below.

\section{The analysis of the Einstein-scalar Lichnerowicz equation}
\label{proofs}

In this section we establish the validity of Tables 1 and 2.
Note first that in stating Theorem \ref{nothm} and Theorem \ref{yesthm}, and in the analysis given below, we do not indicate the degree of regularity needed for the conformal data, nor do we say anything about the regularity of  the solution guaranteed to exist by the relevant theorem. Since optimizing regularity is not one of the central goals or concerns of this paper, we shall generally assume that all sets of conformal data are smooth, and it 
is straightforward to verify that  the solutions produced are then smooth as well.  Our results here easily extend to cases of finite degrees
of regularity, e.g.\ conformal data sets in H\"older spaces $C^{k,\alpha}(\Sigma)$ for $k\geq 2$.  Inspired by low regularity results
for the Einstein evolution equation, a number of existence results for the constraint equations with low regularity assumptions
on the data have been proprovenven in recent years \cite{CB04, Max2, Max3, CBIP, CB06}.  In \cite{CBIP} results similar to those obtained
here are established for the Einstein-Scalar field system on Asymptotically Euclidean manifolds. In that paper the results are
formulated and proved in a low regularity setting.   One could, if desired, obtain similar results for rough Einstein-Scalar field data 
on compact manifolds of the degree of regularity discussed in \cite{CB04, Max2} for the vacuum case. We 
note as well that while the sub and super solution theorem says nothing about uniqueness, 
we can generally obtain uniqueness for solutions of the Lichnerowicz equation using other techniques. (See, e.g., \cite{I95}.) 
 
We have alluded to the conformal covariance of the Lichnerowicz equation a number of times.  While the conformally formulated 
momentum constraint is {\it not} conformally covariant with respect to the splitting of the data we are employing here (see \cite{PY03} for 
an alternative approach) there is an appropriate sense in which the Lichnerowicz equation, including $W$, remains so.
Since we make repeated use of this property, we now state it formally.
\begin{proposition}
\label{confinv}
Let  ${\cal D}=(\g, \sigma, \tau, \psi, \pi)$  be a conformal initial data set for the Einstein-scalar field constraint 
equations on $\Sigma$. If $\tilde\g = \theta^{\frac{4}{n-2}}\g$ for a smooth, positive function $\theta$, 
then we define the corresponding conformally transformed
initial data set by
\begin{equation}
\label{confinitialdata}
{\tilde{\cal D}}=(\tilde\g, \tilde\sigma, \tilde\tau, \tilde\psi, \tilde\pi) = (\theta^{\frac{4}{n-2}} \g, \theta^{-2}\sigma, \tau,  \psi, \theta^{-\frac{2n}{n-2}}\pi).
\end{equation}
Let $W$ be the solution to the conformal form of the momentum constraint equation (\ref{confmom}) 
with respect to the conformal initial data set ${\cal D}$ (for which we assume that a solution exists), and let $\tilde W$ be the solution to (\ref{confmom}) 
with respect to the conformally transformed initial data set ${\tilde{\cal D}}$ (which will exist if $W$ does).  
Then $\phi$ is a solution to the Einstein-scalar field Lichnerowicz 
equation for the conformal data  ${\cal D}$ with $W$
\[
\Delta_{\gamma}\phi -  {\cal R}_{\gamma, \psi}\,\phi + {\cal A}_{\gamma, W, \pi}\,\phi^{-\frac{3n-2}{n-2}}
-  {\cal B}_{\tau, \psi}\,\phi^{\frac{n+2}{n-2}}=0
\]
if and only if $\theta^{-1}\phi$ is a solution to the Einstein-scalar field Lichnerowicz 
equation for the transformed conformal data ${\tilde{\cal D}}$ with $\tilde W$
\[
\Delta_{\tilde\gamma}(\theta^{-1}\phi) -  {\cal R}_{\tilde\gamma, \tilde\psi}\,(\theta^{-1}\phi) + 
{\cal A}_{\tilde\gamma, \tilde W, \tilde\pi}\,(\theta^{-1}\phi)^{-\frac{3n-2}{n-2}}
-  {\cal B}_{\tilde\tau, \tilde\psi}\,(\theta^{-1}\phi)^{\frac{n+2}{n-2}}=0.
\]
\end{proposition}
Note that ${\cal B}_{\tilde\tau, \tilde\psi}= {\cal B}_{\tau, \psi}$; we henceforth use this without further
comment.  The  proof of Proposition \ref{confinv} is left to the reader.

In preparation for the proof of Theorems \ref{nothm} and \ref{yesthm}, we begin by rewriting equation \bref{lich} as
\beq
\label{simplich}
\Delta_{\gamma}\phi =  \mathcal{F}_{\g, \sigma, \tau, \psi, \pi}(\phi)
\eeq
where
\beq
 \mathcal{F}_{\g, \sigma, \tau, \psi, \pi}(\phi)= {\cal R}_{\gamma, \psi}\,\phi - {\cal A}_{\gamma, W, \pi}\,\phi^{-\frac{3n-2}{n-2}}
+  {\cal B}_{\tau, \psi}\,\phi^{\frac{n+2}{n-2}}
\eeq
and the coefficients are defined as in \bref{shortlich}.

 \subsection{Non-existence results}
 
 \noindent
 {\bf Proof of Theorem \ref{nothm}:}
All of the cases in which we assert the non-existence of a positive solution to the Lichnerowicz equation are established
by the following simple integration argument.  Since $\Delta_{\gamma}\phi = \di_\g (\nabla \phi)$, integrating  \bref{simplich} over the compact 
manifold $\Sigma$ yields
\beq
\int_\Sigma \left(\mathcal{F}_{\g, \sigma, \tau, \psi, \pi}(\phi)\right) d\eta_\g= \int_\Sigma \left({\cal R}_{\gamma, \psi}\,\phi - {\cal A}_{\gamma, W, \pi}\,\phi^{-\frac{3n-2}{n-2}} +  {\cal B}_{\tau, \psi}\,\phi^{\frac{n+2}{n-2}}\right) d\eta_\g =0.
\eeq
Thus in the cases in which $\mathcal{F}_{\g, \sigma, \tau, \psi, \pi}(\phi)$ is either greater than or equal  to zero or less than or equal to zero,
but not identically equal to zero, we obtain a contradiction. 
\eop

Note that similar non-existence results have been obtained in \cite{CBIP} for \bref{lich} on non-compact, asymptotically Euclidean manifolds.  
In that setting the integration argument given above is replaced by various forms of the maximum principle  \cite{GT, PW}.

\subsection{Existence results: Proof  of Theorem \ref{yesthm}}
The key tool for proving the existence results asserted in Table 1 and Table 2 is the sub and super solution theorem. 
We state this here in a form particularly suited for our present purposes (and refer to \cite{I95} for a proof).  
\begin{theorem}
\label{subsuper}
If, for the chosen conformal data, there exist a pair of positive functions $\phi_+\geq \phi_-$ such that $\Delta_\lambda \phi_+\leq  \mathcal{F}_{\g, \sigma, \tau, \psi, \pi} (\phi_+)$ and $\Delta_\lambda \phi_-\geq  \mathcal{F}_{\g, \sigma, \tau, \psi, \pi} (\phi_-)$, then there exists a solution $\phi $ of the Einstein-scalar Lichnerowicz equation \bref{simplich} (equivalently \bref{lich}), with $\phi_+\geq \phi \geq \phi_-$.
\end{theorem}

\subsubsection{Existence via constant sub and super solutions}

\noindent
 {\bf Proof of Theorem \ref{yesthm}:}
We begin with the simplest case, namely that in which the Yamabe-scalar field class is zero and the coefficients ${\cal A}_{\gamma, W, \pi}$ and  ${\cal B}_{\tau, \psi}$ are both identically zero.  Note first that ${\cal A}_{\gamma, W, \pi} \equiv 0$ implies $\pi\equiv 0$.  
Referring to \bref{confmom}, this implies (in the CMC setting)
that  $W$ is conformal Killing, $\calD W \equiv 0$ (of course we may have $W\equiv 0$). 
We may then conclude that $\sigma \equiv 0$ as well.  The 
assumption that ${\cal B}_{\tau, \psi}\equiv 0$ and $\tau$ is constant, implies that $V(\psi)$ is constant.  
Regardless of these reductions, if we are working
with a background metric $\g$ for which ${\cal R}_{\gamma, \psi} \equiv 0$ then we see that 
the Lichnerowicz equation reduces to the Laplace equation,
$\Delta_\g \phi =0$.  The relevant solutions are simply the positive constants which correspond to 
the observation that the equality $R(\g) = |\psi|^2_{\g}$ is scale invariant.  

There are two additional classes in which we may prove existence by directly showing that there are constant sub and super solutions. 
Consider first the case in which the metric is in the negative Yamabe-scalar field class, ${\cal A}_{\gamma, W, \pi}$ is identically zero, and  
${\cal B}_{\tau, \psi}$ is
strictly positive.  One then easily verifies that for $\phi_-=\epsilon$, a sufficiently small constant, we have 
$0=\Delta_\g \phi_- \geq \mathcal{F}_{\g, \sigma, \tau, \psi, \pi} (\phi_-)$, so that $\phi_-$ is a  positive subsolution.  Similarly, one sees that 
 $\phi_+=\epsilon^{-1}$, for $\epsilon$  sufficiently small, is a positive supersolution.  Since $\phi_-< \phi_+$, Theorem \ref{subsuper} 
 applies and produces a positive solution to \bref{lich}.
One may easily check that the same approach also works when ${\cal A}_{\gamma, W, \pi}$ is not  identically zero.

In order to establish the other affirmative answers for the other cases of initial data, we will use the method presented
by   Maxwell in \cite{Max2} for the vacuum setting. 
\begin{proposition}
\label{maxprop1}
There exists a positive solution of the Einstein-scalar Lichnerowicz equation provided either
\begin{itemize}
\item[(i)] ${\mathcal Y}_{\psi}([\gamma])\geq0$, ${\cal A}_{\gamma, W, \pi}\not\equiv 0$ and ${\cal B}_{\tau, \psi} >0$ or
${\cal B}_{\tau, \psi}\geq 0$
\item[(ii)] ${\mathcal Y}_{\psi}([\gamma])>0$ ${\cal A}_{\gamma, W, \pi}\not\equiv 0$ and ${\cal B}_{\tau, \psi}\equiv 0$
\end{itemize}
\end{proposition}
{\bf Proof:} We begin by assuming, via Proposition \ref{tfae} and Proposition \ref{confinv}, 
that we have chosen  a background metric $\g$ and
associated conformal data so that 
\[
\mathrm{sign}({\cal R}_{\gamma, \psi}) = \mathrm{sign}({\mathcal Y}_{\psi}([\gamma])).
\]
Under our hypothesis  we know that $({\cal R}_{\gamma, \psi} + {\cal B}_{\tau, \psi})$ is non-negative and not identically zero;
therefore there is a solution $\phi_2$ to
\begin{equation}
\label{lin1}
-\Delta_\g \phi_2 + ({\cal R}_{\gamma, \psi} + {\cal B}_{\tau, \psi})\phi_2 = {\cal A}_{\gamma, W, \pi}.
\end{equation}
Since ${\cal A}_{\gamma, W, \pi}\not\equiv 0$ and ${\cal A}_{\gamma, W, \pi}\geq 0$, by the maximum principle we have 
$\phi_2>0$.  Setting ${\tilde\g} = \phi_2^{\frac{4}{n-2}}\g$ and transforming the other pieces of the conformal data as in
Proposition \ref{confinv}, we
observe that the existence of a positive solution of the Einstein-scalar Lichnerowicz equation is equivalent to the existence
of a positive solution to 
\begin{equation}
\label{newlich}
\Delta_{\tilde\g} \phi =  {\cal R}_{\tilde\gamma, \tilde\psi}\phi - {\cal A}_{\tilde\gamma, \tilde W, \tilde\pi}\phi^{-\frac{3n-2}{n-2}}+
 {\cal B}_{\tilde\tau, \tilde\psi}\phi^{\frac{n+2}{n-2}}.
\end{equation}
We compute using (\ref{lin1})
\begin{eqnarray*}
{\cal R}_{\tilde\gamma, \tilde\psi} &=& c_n (R(\tilde{\gamma}) - |\nabla \psi|_{\tilde{\gamma}}^2)\\
 &=& - \phi^{-\frac{n+2}{n-2}} (\Delta_\g \phi_2 - {\cal R}_{\gamma, \psi} \phi_2)\\
 &=& \phi^{-\frac{n+2}{n-2}} ({\cal A}_{\gamma, W, \pi} - {\cal B}_{\tau, \psi}\phi_2 )\\
 &=& \phi^{-\frac{n+2}{n-2}} (\phi_2^{\frac{4n}{n-2}}{\tilde{\cal A}}_{\gamma, W, \pi} - {\cal B}_{\tau, \psi}\phi_2 )\\
  &=&{\cal A}_{\tilde\gamma, \tilde W, \tilde \pi} \phi_2^{\frac{3n-2}{n-2}} - {\cal B}_{\tau, \psi}\phi_2^{-\frac{4}{n-2}}.
\end{eqnarray*}
Using this we see that (\ref{newlich}) becomes
\begin{equation}
\label{newlich2}
\Delta_{\tilde\g} \phi =  {\cal A}_{\tilde\gamma, \tilde W, \tilde \pi}(\phi_2^{\frac{3n-2}{n-2}}\phi-\phi^{-\frac{3n-2}{n-2}}) -
 {\cal B}_{\tau, \psi}(\phi_2^{-\frac{4}{n-2}}\phi-\phi^{\frac{n+2}{n-2}}).
\end{equation}
A constant $\phi_+$ is a super solution of (\ref{newlich2}) if
\begin{equation}
\label{reqineq1}
{\cal A}_{\tilde\gamma, \tilde W, \tilde \pi}\phi_2^{\frac{3n-2}{n-2}}-{\cal B}_{\tau, \psi}\phi_2^{-\frac{4}{n-2}}\geq
{\cal A}_{\tilde\gamma, \tilde W, \tilde \pi}\phi_+^{-\frac{4(n-1)}{n-2}} -{\cal B}_{\tau, \psi} \phi_+^{\frac{4}{n-2}}.
\end{equation}
Pick $\phi_+$ such that 
\[
\phi_+^{\frac{4(n-1)}{n-2}}\geq \sup_{\Sigma} \phi_2^{-\frac{3n-2}{n-2}} \qquad{\mbox{and}}\qquad
\phi_+^{\frac{4}{n-2}}\geq \sup_{\Sigma} \phi_2^{-\frac{4}{n-2}}.
\]
Then 
\[
{\cal A}_{\tilde\gamma, \tilde W, \tilde\pi}\phi_2^{\frac{3n-2}{n-2}}\geq{\cal A}_{\tilde\gamma, \tilde W, \tilde\pi}\phi_+^{-\frac{4(n-1)}{n-2}}
\qquad{\mbox{and}}\qquad
- {\cal B}_{\tau, \psi}\phi_2^{-\frac{4}{n-2}}\geq- {\cal B}_{\tau, \psi}\phi_+^{\frac{4}{n-2}}
\]
so (\ref{reqineq1}) is satisfied.  A similar argument shows that a positive constant $\phi_-$ which satisfies
\[
\phi_-^{\frac{4(n-1)}{n-2}}\leq \inf_{\Sigma} \phi_2^{-\frac{3n-2}{n-2}} \qquad{\mbox{and}}\qquad
\phi_-^{\frac{4}{n-2}}\leq \inf_{\Sigma} \phi_2^{-\frac{4}{n-2}}
\]
is a sub solution of   (\ref{newlich2}).  Moreover $\phi_-\leq \phi_+$, so we may apply Theorem \ref{subsuper} to conclude
the existence of a positive solution to  (\ref{newlich2}) and therefore to the original Einstein-scalar field Lichnerowicz 
equation.
This completes the proof of Proposition \ref{maxprop1} and thus Theorem \ref{yesthm}. \eop

\subsection{Partial results: reduction to previously known results for ${\mathcal Y}_{\psi}([\gamma])<0$ and ${\cal B}_{\tau, \psi}\geq 0$}
\label{PR}
There are two cases with a negative Yamabe-scalar field conformal invariant for which we can obtain partial results. 
The situation here is somewhat different from the partial results described in the next section in that we indicate
the possibility of finding necessary and sufficient conditions for the existence of a positive
solution to the Einstein-scalar field Lichnerowicz equation in these cases.  

We begin by showing that the case in which ${\cal A}_{\gamma, W, \pi}\not\equiv 0$ can be reduced to the 
${\cal A}_{\gamma, W, \pi}\equiv 0$ case; then the problem is simply one of prescribed ${\cal R}_{\gamma, \psi}$.
\begin{proposition}
\label{redpsp}
Suppose that ${\mathcal Y}_{\psi}([\gamma])<0$, ${\cal A}_{\gamma, W, \pi}\not\equiv 0$ and ${\cal B}_{\tau, \psi}\geq 0$.
Then the Einstein-scalar field Lichnerowicz equation has a positive solution if and only if there is a $\phi_1>0$ such that 
$\tilde\g = \phi_1^{\frac{4}{n-2}}\g$ satisfies
\[
{\cal R}_{\tilde\gamma, \tilde\psi}= - {\cal B}_{\tau, \psi}. 
\]
\end{proposition}
{\bf Proof:}
We again follow the argument given by Maxwell  \cite{Max2} in the vacuum case. 
We first suppose that there exists such a  $\phi_1>0$. Since ${\mathcal Y}_{\psi}([\gamma])<0$, we deduce that 
 ${\cal B}_{\tau, \psi}\not\equiv 0$. The Einstein-scalar field Lichnerowicz equation is equivalent to 
\begin{equation}
\label{newlich3}
\Delta_{\tilde\g} \phi  +{\cal B}_{\tau, \psi}\phi ={\cal A}_{\tilde\gamma, \tilde W, \tilde\pi}\phi^{-\frac{3n-2}{n-2}}-
 {\cal B}_{\tau, \psi}\phi^{\frac{n+2}{n-2}}.
\end{equation}
Since  ${\cal B}_{\tau, \psi}\not\equiv 0$ and  ${\cal A}_{\gamma, W, \pi}\not\equiv 0$ there exists a unique
positive solution, $\phi_2$, of 
\begin{equation}
\label{lineq2}
-\Delta_{\tilde\g} \phi_2  +{\cal B}_{\tau, \psi}\phi_2 ={\cal A}_{\tilde\gamma, \tilde W, \tilde\pi}.
\end{equation}
Set $\hat\g = \phi_2^{\frac{4}{n-2}} \tilde\g$ and let $(\hat\sigma, \hat\tau, \hat W, \hat \psi, \hat\pi)$ be the rest
of the conformally transformed initial data set.  We compute, using the assumption that 
${\cal R}_{\tilde\gamma, \tilde\psi}= - {\cal B}_{\tau, \psi}$  and (\ref{lineq2}),
\begin{eqnarray*}
{\cal R}_{\hat\gamma, \hat\psi} &=& \phi_2^{-\frac{n+2}{n-2}} (-\Delta_{\tilde\g} \phi_2  -{\cal B}_{\tau, \psi}\phi_2)\\
&=&\phi_2^{-\frac{n+2}{n-2}} (-2{\cal B}_{\tau, \psi}\phi_2 + {\cal A}_{\tilde\gamma, \tilde W, \tilde\pi})\\
&=&-2{\cal B}_{\tau, \psi}\phi_2^{-\frac{4}{n-2}} + {\cal A}_{\hat\gamma, \hat W, \hat\pi}\phi_2^{\frac{3n-2}{n-2}}.
\end{eqnarray*}
So the Einstein-scalar field Lichnerowicz equation with respect to $(\hat\g, \hat\sigma, \hat\tau, \hat W, \hat \psi, \hat\pi)$ is
\begin{eqnarray*}
\Delta_{\hat\g}\phi &=& {\cal R}_{\hat\gamma, \hat\psi} \phi -  {\cal A}_{\hat\gamma, \hat W, \hat\pi}\phi^{-\frac{3n-2}{n-2}}
+{\cal B}_{\tau, \psi}\phi_2^{\frac{n+2}{n-2}}\\
&=& -2{\cal B}_{\tau, \psi}\phi_2^{-\frac{4}{n-2}}\phi + {\cal A}_{\hat\gamma, \hat W, \hat\pi}\phi_2^{\frac{3n-2}{n-2}}\phi
-  {\cal A}_{\hat\gamma, \hat W, \hat\pi}\phi^{-\frac{3n-2}{n-2}} +{\cal B}_{\tau, \psi}\phi_2^{\frac{n+2}{n-2}}.
\end{eqnarray*}
As before we may now verify the existence of positive, constant sub and super solutions for this equation.  Let $\phi_+$ 
satisfy 
\[
\phi_+^{\frac{3n-2}{n-2} +1} \geq \sup_{\Sigma} \phi_2^{-\frac{3n-2}{n-2}}
\qquad \mbox{and}\qquad
\phi_+^{\frac{4}{n-2}} \geq 2\sup_{\Sigma} \phi_2^{-\frac{4}{n-2}}.
\]
Then one may verify that $\phi_+$ is a positive, constant super solution. Similarly, let $\phi_-$ 
satisfy 
\[
\phi_-^{\frac{3n-2}{n-2} +1} \leq \inf_{\Sigma} \phi_2^{-\frac{3n-2}{n-2}}
\qquad \mbox{and}\qquad
\phi_-^{\frac{4}{n-2}} \leq 2\inf_{\Sigma} \phi_2^{-\frac{4}{n-2}}.
\]
Then one may verify that $\phi_-$ is a positive, constant sub solution.  Moreover, $\phi_-\leq \phi_+$, so we conclude that there
exists a positive solution to  the Einstein-scalar field Lichnerowicz equation with respect to 
$(\hat\g, \hat\sigma, \hat\tau, \hat W, \hat \psi, \hat\pi)$ and therefore, by Proposition \ref{confinv}, to our original equation. 
This establishes the desired implication.

We now suppose that $\g$ satisfies ${\cal R}_{\gamma, \psi}<0$ and that the Einstein-scalar field Lichnerowicz equation with respect to 
$(\g, \sigma, \tau,  W,  \psi, \pi)$ admits a positive solution.  We wish to solve
\begin{equation}
\label{dlin}
\Delta_\g \phi - {\cal R}_{\gamma, \psi} \phi = {\cal B}_{\tau, \psi} \phi^{\frac{n+2}{n-2}}
\end{equation}
so that $\tilde\g =  \phi^{\frac{4}{n-2}}\g$ satisfies ${\cal R}_{\tilde\gamma, \tilde\psi} = -{\cal B}_{\tau, \psi}$.
We denote the positive solution of the  Einstein-scalar field Lichnerowicz equation by $\phi_+$.
It holds that:
\begin{eqnarray*}
\Delta_\g \phi_+ - {\cal R}_{\gamma, \psi} \phi_+ &=& -{\cal A}_{\gamma,  W, \pi}\phi^{-\frac{3n-2}{n-2}}+ 
{\cal B}_{\tau, \psi} \phi_+^{\frac{n+2}{n-2}}\\
&\leq&{\cal B}_{\tau, \psi} \phi_+^{\frac{n+2}{n-2}}.
\end{eqnarray*}
Thus $\phi_+$ is a positive super solution of (\ref{dlin}). In order to find a positive sub solution we proceed as follows.
Consider
\begin{equation}
\label{eeq}
\Delta_\g \phi_\epsilon + {\cal R}_{\gamma, \psi} \phi_\epsilon = {\cal R}_{\gamma, \psi} +\epsilon {\cal B}_{\tau, \psi}.
\end{equation}
Since ${\cal R}_{\gamma, \psi}<0$, (\ref{eeq}) has a unique solution for each $\epsilon$.  When $\epsilon =0$ the solution is
$\phi_0\equiv 1$.  Therefore, for $\epsilon$ sufficiently small, we may ensure that $\phi_\epsilon > \frac12$. We claim that 
$\phi_- = \eta\phi_{\epsilon}$ for $\eta$ sufficiently small is an appropriate positive sub solution.  First pick $\eta$ so that 
\[
\eta\phi_{\epsilon} < \phi_+ \qquad \mbox{and} \qquad \eta^{\frac{4}{n-2}} < \epsilon\inf_{\Sigma} \phi_\epsilon^{-\frac{n+2}{n-2}}.
\]
Then
\begin{eqnarray*}
\Delta_\g \phi_- - {\cal R}_{\gamma, \psi} \phi_- &=& \eta({\cal R}_{\gamma, \psi}(1-2\phi_\epsilon) +\epsilon {\cal B}_{\tau, \psi})\\
&\geq& \eta\,\epsilon\, {\cal B}_{\tau, \psi}\\
&\geq& \eta(\eta^{\frac{4}{n-2}} \phi_\epsilon^{\frac{n+2}{n-2}}){\cal B}_{\tau, \psi}\\
&=&\phi_-^{\frac{n+2}{n-2}}{\cal B}_{\tau, \psi}.
\end{eqnarray*}
So $\phi_-$ is a positive subsolution with $0<\phi_-<\phi_+$.  Therefore there exists a positive solution $\phi$ of (\ref{dlin}).
This completes the proof of Proposition \ref{redpsp}. 
\eop

We have now reduced the problem of solving the Einstein-scalar field Lichnerowicz equation for initial data with 
a negative Yamabe-scalar field conformal invariant and  ${\cal B}_{\tau, \psi}\geq 0$ to the prescribed
scalar curvature-scalar field problem, namely the existence of a conformally related metric
$\tilde\g = \phi^{\frac{4}{n-2}}\g$ which satisfies
\begin{equation}
\label{pscsf}
{\cal R}_{\tilde\gamma, \tilde\psi}= - {\cal B}_{\tau, \psi}. 
\end{equation}
In the special case that $\tau=0$, $\psi\equiv 0$ and $4V(\psi)=-F$, this reduces to the prescribed scalar curvature problem
\begin{equation}
\label{psc}
 R(\tilde\gamma) = -F 
\end{equation}
with $F\geq 0$ but not strictly positive.
A. Rauzy \cite{Rau} has provided necessary and sufficient conditions for the existence of solution to this problem.
In order to state his condition (using the formulation given in \cite{Max2}) we proceed as follows.  Since we are
in a negative Yamabe conformal class, we may first make a conformal transformation to a metric, which we 
again denote by $\g$, with constant negative scalar curvature $R(\g)$ and such that $\mathrm{Vol}_\g(\Sigma)=1$.
We use the  prescribed function $-F$ to define a subset of $H^1(\Sigma)$:
\[
\mathscr{A} = \{ f\in H^1(\Sigma): f\geq 0, f\not\equiv 0 \mbox{ and } \int_\Sigma f\, F dv_{\gamma} =0\}.
\]
Then there is a $\tilde\g = \phi^{\frac{4}{n-2}}\g$ which satisfies (\ref{psc}) if and only if
\begin{equation}
\label{rauzy}
\inf_{f\in \mathscr{A}} \frac{\int_\Sigma c_{n}^{-1} |\nabla f|_\g^2 dv_\g}{\int_\Sigma  f^2 dv_\g} > -R(\g).
\end{equation}
We conjecture that there is a natural extension  of Rauzy's condition (\ref{rauzy})  to the scalar curvature-scalar field setting 
which also provides for necessary and sufficient conditions for the existence of a conformal metric satisfying (\ref{pscsf}).  
We remark that for either of these problems, it is generally quite difficult to determine, for a given conformal class and prescribed function
(or set of conformal data) whether  Rauzy's condition (or its extension) is actually satisfied.

\subsection{Partial results: reduction to previously known results for ${\mathcal Y}_{\psi}([\gamma])>0$}
\label{PR2}
Two exceptional cases occur if ${\cal A}_{\gamma, W, \pi}\equiv 0$ and we are in the positive Yamabe-scalar field class with either
${\cal B}_{\tau, \psi}$ is strictly negative, or less than or equal to zero, but not identically zero, on $\Sigma$.  In this 
case we may rewrite \bref{simplich} as 
\beq
\label{pSCSFeqn}
\Delta_{\gamma}\phi =   {\cal R}_{\gamma, \psi}\,\phi +  {\cal B}_{\tau, \psi}\,\phi^{\frac{n+2}{n-2}}.
\eeq
If $\phi$ is a positive solution to \bref{pSCSFeqn} then the metric $\tilde\g = \phi^{\frac{4}{n-2}}\g$ satisfies
\[
R(\tilde\g) - |\nabla\psi|^2_{\tilde\g} = -{\cal B}_{\tau, \psi}\geq 0.
\]
If $\nabla\psi \equiv 0$ and if ${\cal B}_{\tau, \psi}$ is a strictly negative constant 
(This occurs if $\tau$, $\psi$ and $V(\psi)$ are constant and $V(\psi)> \frac{(n-1)}{4n} \tau^2$.)
this becomes
the question of the existence of a metric of constant positive scalar curvature within a positive Yamabe class, i.e. the difficult 
case of the Yamabe problem.  By appealing to the solution to the Yamabe problem, and possibly rescaling to obtain the desired 
constant, we may assert the existence of a solution in this case.  Other examples of existence in these cases may be obtained 
by appealing to results on the prescribed scalar curvature problem (see \cite{Aub98} for a partial survey). 
For example Escobar and Schoen have shown
that if $\gamma$ is not in the conformal class of the standard metric on the round sphere in three dimensions, 
any function $f= -{\cal B}_{\tau, \psi}$
is the scalar curvature of a conformally related metric provided $\sup f >0$ \cite{ES}.  We note that 
we are not establishing any  new results in these cases, but rather pointing out the relevance of these known results to special 
cases of the problem at hand.  We expect that many of the known results for the  prescribed scalar curvature problem may be
easily extended to give results for this class of initial data if $\nabla\psi \not\equiv 0$ and if $V$ is appropriately chosen.

\subsection{Partial results:  new results when ${\cal B}_{\tau, \psi}<0$}
\label{FPR}
We now consider the class of conformal data which have a positive Yamabe-scalar field conformal invariant, have 
${\cal A}_{\gamma, W, \pi}$ not  identically zero and have ${\cal B}_{\tau, \psi}$ strictly negative. We have not to date determined 
whether there exists a solution for every set of data in this class.  However, we have determined a condition on the data which is 
sufficient for the existence of a solution.  A similar situation was previously treated in \cite{DN} in the study of initial data for fluid bodies.
They were also able to obtain partial existence results (see Theorem 2 of \cite{DN}).

Our results for this class rely on verifying, for certain choices of the conformal data $(\g, \sigma, \tau, \psi, \pi)$ in this class,
that the Lichnerowicz equation admits constant positive sub and super solutions, and therefore admits a positive solution.
To verify the existence of these constant positive sub and super solutions it is sufficient to show that if we make a choice 
of $\g$ in the conformal class for which ${\cal R}_{\gamma, \psi}>0$, then there exists constant 
$C_+\geq C_- >0$ such that for all $x\in \Sigma$, 
$\mathcal{F}_{\g, \sigma, \tau, \psi, \pi}(C_+)\geq 0$ and $\mathcal{F}_{\g, \sigma, \tau, \psi, \pi}(C_-)\leq 0$.
Multiplying $\mathcal{F}_{\g, \sigma, \tau, \psi, \pi}(\phi)$ by $\phi^{\frac{3n-2}{n-2}}$ and setting 
$y= \phi^{\frac{4}{n-2}}$,  we see that it is sufficient to find constants $m\geq \ell>0$ for which $h(x, m) \geq 0$
and $h(x, \ell) \leq 0$ for all $x\in \Sigma$, where
\[
h(x,y) = {\cal B}_{\tau, \psi}(x)y^n + {\cal R}_{\gamma, \psi}(x)y^{n-1} - {\cal A}_{\gamma, W, \pi}(x),
\]
with ${\cal B}_{\tau, \psi}(x)$ strictly negative and ${\cal R}_{\gamma, \psi}(x)$ strictly positive.
If we assume that ${\cal A}_{\gamma, W, \pi}(x)$ is non-zero for all $x\in \Sigma$, we see that 
$h(x, 0) = - {\cal A}_{\gamma, W, \pi}(x)<0$; it follows that there exists arbitrarily small, positive $y$ for which $h(x, 0)<0$.
Now examining the behavior of $h(x,y)$ as a function of $y\geq 0$ for fixed $x$, we see that $h$ increases from 
$h(x, 0)= - {\cal A}_{\gamma, W, \pi}(x)$ to a maximum at 
$y_0 (x) = -\frac{n-1}{n} \frac{{\cal R}_{\gamma, \psi}(x)}{{\cal B}_{\tau, \psi}(x)}>0$
and then monotonically decreases (to $-\infty$) for all $y> y_0$. Hence there exists a constant, positive super solution
$m$ only if for each $x$,  $h(x, y_0(x))>0$.  We readily verify that $h(x, y_0(x))>0$ so long as 
\[
\left(\frac{n-1}{n}\right)^{n-1} \left(\frac1n\right) \frac{({\cal R}_{\gamma, \psi}(x))^n}{(- {\cal B}_{\tau, \psi}(x))^{n-1}} > {\cal A}_{\gamma, W, \pi}(x).
\]
It follows that if 
\begin{equation}
\label{suffcond}
\left(\frac{n-1}{n}\right)^{n-1} \left(\frac1n\right) \frac{\inf_\Sigma ({\cal R}_{\gamma, \psi})^n}{\sup_\Sigma(- {\cal B}_{\tau, \psi})^{n-1}} 
>\sup_{\Sigma} {\cal A}_{\gamma, W, \pi}.
\end{equation}
then indeed $h(x, y_0(x))>0$ for all $x\in \Sigma$.

If in fact $h(x, y_0(x))>0$, then from continuity we deduce that for each $x\in \Sigma$ there is an interval $I(x)$ such that for 
$y\in I(x)$, $h(x, y)>0$.  We therefore see that a constant super solution exists so long as $\cap_{x\in \Sigma} I(x)$ is nonempty.
We have proven  the following:
\begin{theorem} Assume that we are given a compact manifold $\Sigma$, and conformal data 
$(\g, \sigma, \tau, \psi, \pi)$ on $\Sigma$, with $\tau$ constant, and with the data satisfying the conditions
\begin{itemize}
\item[(i)] ${\cal R}_{\gamma, \psi} >0$ 
\item[(ii)] ${\cal B}_{\tau, \psi}<0$
\item[(iii)] ${\cal A}_{\gamma, W, \pi}>0$
\item[(iv)] inequality (\ref{suffcond}) is valid
\item[(v)] $\cap_{x\in \Sigma} I(x)$ is nonempty.
\end{itemize}
If either $(\Sigma, \g)$ admits no conformal Killing vector fields, or $\pi\nabla\psi$ is orthogonal 
to the space of conformal Killing vector fields, then we may
find a positive function $\phi$ and a vector field $W$ so that the conformal reconstructed data 
$({\bar \g}, {\bar K}, {\bar \psi}, {\bar \pi})$, defined in equations \bref{confmetric}, \bref{confK}, \bref{psi} and \bref{pi},   
satisfy the Einstein-scalar field constraint equations \bref{ham}-\bref{mom}.  In the case that there are no conformal 
Killing fields, the resulting solution is unique.
\end{theorem}

\subsection{Partial results: when ${\cal B}_{\tau, \psi}$ changes sign}
\label{YCB}
In this section we consider conformal data for which  ${\cal B}_{\tau, \psi}=c_n\left(\frac{n-1}{n} \tau^2 -4V(\psi) \right)$
takes both positive and negative values on the manifold $\Sigma$.  This can occur if $\tau$ is non-zero and the 
scalar potential $V(\psi)$ is a positive function.  The results we have obtained thus far require that the 
Yamabe-scalar field invariant  ${\mathcal Y}_{\psi}([\gamma])$ be positive; see \cite{CB06} for a discussion of the 
other Yamabe-scalar field classes. Thus we presume, after possibly making an initial conformal modification of our data,
that ${\cal R}_{\gamma, \psi}>0$.

We seek conditions on the conformal data sufficient to guarantee that the Lichnerowicz equation has a constant
subsolution $\phi_-$ and a constant supersolution $\phi_+$.  Using the notation of section \ref{FPR}, we find it sufficient 
that there exist constants  $m\geq \ell >0$ such that
\beq
h(x,\ell )\equiv {\cal B}_{\tau, \psi}(x)\ell^{n}+{\cal R}_{\gamma, \psi}(x)\ell ^{n-1}-{\cal A}_{\gamma, W, \pi}(x)\leq 0
\label{subineq}
\eeq
\beq
h(x,m)\equiv {\cal B}_{\tau, \psi}(x)m^{n}+{\cal R}_{\gamma, \psi}(x)m^{n-1}-{\cal A}_{\gamma, W, \pi}(x)\geq 0
\label{supineq}
\eeq
at each point $x\in\Sigma$. It follows that $\phi_+ = m^{\frac{n-2}{4}}$ and $\phi_- = \ell^{\frac{n-2}{4}}$ are then 
constant sub and supersolutions.

We analyze separately the regions of $\Sigma$ in which ${\cal B}_{\tau, \psi}$ is nonnegative,
labeled $\Sigma_{+}$  and where ${\cal B}_{\tau, \psi}$ is non-positive, labeled $\Sigma_{-}$ 
(so that $\Sigma = \Sigma_{+}\cup \Sigma_{-}$).

\medskip

\noindent {\bf (a)} In $\Sigma_{+} :=\{x\in\Sigma | {\cal B}_{\tau, \psi}(x)\geq 0\} $.

\smallskip

We use the trivial inequalities ${\cal R}_{\gamma, \psi}m^{n-1}\geq 
{\cal R}_{\gamma, \psi}$ and ${\cal A}_{\gamma, W, \pi}m^{\frac{n-1}{2}}\geq {\cal A}_{\gamma, W, \pi}$ which
(since ${\cal A}_{\gamma, W, \pi}\geq 0$ and ${\cal R}_{\gamma, \psi}\geq 0$ by assumption) hold for
$m\geq 1$,  to note that sufficient
conditions for $\ell _{+}$ and $m_{+}$ to satisfy the
inequalities  (\ref{subineq})--(\ref{supineq}) in $\Sigma_{+}$ are given by
\begin{equation}
m_{+}^{\frac{n-1}{2}}\geq \max\{1,\frac{\underset{\Sigma_{+}}{\sup}({\cal A}_{\gamma, W, \pi})}{\underset{\Sigma_{+}}{\inf}({\cal B}_{\tau, \psi}+{\cal R}_{\gamma, \psi})}\},\qquad
0<\ell _{+}\leq \min\{1,\frac{\underset{\Sigma_{+}}{\inf}({\cal A}_{\gamma, W, \pi})}{\underset{\Sigma_{+}}
{\sup}({\cal B}_{\tau, \psi}+{\cal R}_{\gamma, \psi})}\}.
\label{splus-conds}
\end{equation}
\medskip

\noindent {\bf (b)} In $\Sigma_{-} :=\{x\in\Sigma | {\cal B}_{\tau, \psi}(x)\leq 0\} $.

\smallskip

Since the function $h(x,y)$ is continuous at $y=0$, and since  $h(x,0)=-{\cal A}_{\gamma, W, \pi}(x)$, 
it follows that there exists a positive constant $\ell_-$ satisfying (\ref{subineq}) on $\Sigma_{-}$ so long as 
\begin{equation}
\underset{\Sigma_{-}}{\inf}{\cal A}_{\gamma, W, \pi}>0.
\label{start}
\end{equation}
One readily calculates that, for each fixed $x$ in $\Sigma_-$, the function 
$h(x,y)$ achieves a maximum  at 
\[
y_{\max}(x)=\frac{(n-1){\cal R}_{\gamma, \psi}(x)}{n| {\cal B}_{\tau, \psi}(x)|}.
\]
and further one has
\[
h(x,y_{\max}(x))= \frac{1}{n}\left[\frac{(n-1)}{n}\right]^{n-1}\frac{{\cal R}_{\gamma, \psi}(x)^{n}}{|%
{\cal B}_{\tau, \psi}(x)|^{n-1}}-{\cal A}_{\gamma, W, \pi}(x).
\]
Therefore the maximum $h(x,y_{\max})$ is positive if and only
\begin{equation}
n\left[\frac{n}{n-1}\right]^{n-1}|{\cal B}_{\tau, \psi}(x)|^{n-1}{\cal A}_{\gamma, W, \pi}(x)<{\cal R}_{\gamma, \psi}(x)^{n}.
\label{roots}
\end{equation}
\begin{remark}
For  $n=3$, condition (\ref{roots}) corresponds to the well known condition for
the 3rd order polynomial $h$  to have 3 real roots:
\[
27({\cal B}_{\tau, \psi}(x))^{2}{\cal A}_{\gamma, W, \pi}(x)<4{\cal R}_{\gamma, \psi}^{3}(x).
\]
\end{remark}

Now if $h(x, y_{\max}(x))>0$, it follows from the form of the function $h(x,y)$ and from the negativity of 
$-{\cal A}_{\gamma, W, \pi}(x)$ (by assumption  (\ref{start})) and of $({\cal B}_{\tau, \psi}(x)$ (since $x\in \Sigma_-$)
that $h(x,y)$ has a pair of positive roots $0<z_{1}(x)\leq $ $z_{2}(x)$. The numbers 
$X(x)$ and $Y(x)$ then satisfy the inequalities (\ref{subineq}) and  (\ref{supineq}) at $x$ so long as
\begin{equation}
0<X(x)\leq z_{1}(x)\leq Y(x)\leq z_{2}(x).\text{\ }
\end{equation}
Therefore, there exist constant numbers $\ell _{-}$ and $m_{-}$ satisfying (\ref{subineq}) and  (\ref{supineq})
on $\Sigma_{-}$ if, in addition to  (\ref{start}) and (\ref{roots}), we have
\begin{equation}
\max_{\Sigma_{-}} z_{1}(x)\leq \min_ {\Sigma_{-}} z_{2}(x).
\label{sneq-cond}
\end{equation}
This leads to an existence theorem for the corresponding 
Lichnerowicz equation, and as a result leads to the following existence theorem for the Einstein-scalar field constraint 
equations, which is valid in the case of a positive 
Yamabe-scalar field class  but which does not assume that ${\cal B}_{\tau, \psi}$ has a fixed sign on $\Sigma$.
\begin{theorem}
Assume that we are given a compact manifold $\Sigma$, and conformal data 
$(\g, \sigma, \tau, \psi, \pi)$ on $\Sigma$, with $\tau$ constant, ${\mathcal Y}_{\psi}([\gamma])>0$, and with the data satisfying the
following
\begin{itemize}
\item[(i)] ${\cal R}_{\gamma, \psi} >0$ 
\item[(ii)] ${\cal A}_{\gamma, W, \pi}>0$
\item[(iii)]  conditions (\ref{splus-conds}), (\ref{roots}), and (\ref{sneq-cond}) for the existence of 
$\ell _{+},\ell _{-}$, $m_{+}$ and $m_{-}$ are satisfied 
\item[(iv)] $\max\{\ell _{+}, \ell _{-}\}\leq \min\{m_{+}, m_{-}\}$.
\end{itemize}
If either $(\Sigma, \g)$ admits no conformal Killing vector fields, or $\pi\nabla\psi$ is orthogonal 
to the space of conformal Killing vector fields, then we may
find a positive function $\phi$ and a vector field $W$ so that the conformal reconstructed data 
$({\bar \g}, {\bar K}, {\bar \psi}, {\bar \pi})$, defined in equations \bref{confmetric}, \bref{confK}, \bref{psi} and \bref{pi},   
satisfy the Einstein-scalar field constraint equations \bref{ham}-\bref{mom}.  In the case that there are no conformal 
Killing fields, the resulting solution is unique.
\label{varsignthm}
\end{theorem}
Theorem \ref{varsignthm} is not optimal. It only gives sufficient conditions
\footnote{It is very probable, in particular, that the
condition ${\cal A}_{\gamma, W, \pi}>0$ on $\Sigma_{-}$ can be replaced by 
 ${\cal A}_{\gamma, W, \pi}\geq 0$.} for the existence of a solution,
and leaves space for further research if physical problems motivate it.
Note that for ${\mathcal Y}_{\psi}([\gamma])>0$ and  ${\cal A}_{\gamma, W, \pi}\not\equiv 0$, this theorem 
provides partial results for the case ${\cal B}_{\tau, \psi}\geq 0$ but not strictly positive, as well as for the 
case when ${\cal B}_{\tau, \psi}$ properly changes sign.

\section{The Einstein-scalar field constraint equations for asymptotically hyperboloidal initial data}
\label{AH}

Having discussed solutions of the Einstein-scalar constraint equations on closed manifolds here, and those which are asymptotically Euclidean in \cite{CBIP}, we wish to comment briefly on the remaining category of interest: initial data solutions which are asymptotically hyperbolic (or  ``hyperboloidal''  \cite{F}). 
Intuitively speaking, a set of initial data $(\Sigma, \gamma, K, \psi, \pi)$, with $\Sigma$ noncompact, 
is hyperboloidal if , as one approaches each 
connected component of infinity in $\Sigma$, the metric $\g$ approaches a metric of constant negative curvature, $K$ approaches a pure 
(non-zero constant) trace tensor, and both $\psi$ and $\pi$ approach zero. Such behavior corresponds to that of initial data induced on a spacelike hypersurface which asymptotically approaches null infinity in an appropriate way in an asymptotically flat spacetime. One may define hyperboloidal data precisely via a conformal compactification of the geometry in terms of a defining function, and via the use of weighted function spaces (see \cite{A, AC, ACF, IMP1}).

We rely upon two key lemmas: The first \cite{IP} shows that the sub and super solution method holds  on hyperboloidal geometries for equations of the Lichnerowicz type. The second \cite{ACF, AM} shows that any hyperboloidal geometry is conformally related to one with constant negative scalar curvature. It follows from this second result that for hyperboloidal data, the function
${\cal R}_{\gamma, \psi}=c_n\left(R(\gamma)-|\nabla{\psi}|^2_{\gamma}\right)$ can always be conformally deformed to a strictly
negative one.

We now see that,  in determining which sets of hyperboloidal conformal data map to hyperboloidal solutions of the Einstein-scalar field constraint equations, we are effectively working with the first row of the Tables 1 and 2.  In particular, it is very easy to show that  if the scalar potential function $V(\psi)$ is such that ${\cal B}_{\tau, \psi}$ is non positive or zero, then the data is not mapped to a solution, while a strictly positive  ${\cal B}_{\tau, \psi}$ guarantees (via constant sub and super solutions) that the data is mapped to a solution. For non negative 
${\cal B}_{\tau, \psi}$, there are likely partial results to be found as well, but we leave these to be determined by the interested reader.

\include{bib}

 \end{document}
